# O-Net: A Convolutional Neural Network for Quantitative Photoacoustic Image Segmentation and Oximetry

Geoffrey P. Luke,[1,2,†,*] Kevin Hoffer-Hawlik,[1,†] Austin C. Van Namen[1], Ruibo Shang[1]

[1]*Thayer School of Engineering, Dartmouth College, Hanover, NH 03755, USA*
[2]*Translational Engineering in Cancer Program, Norris Cotton Cancer Center, Lebanon, NH 03756, USA*
[†]*Equal contribution*
*\*Corresponding author: geoffrey.p.luke@dartmouth.edu*

**Estimation of blood oxygenation with spectroscopic photoacoustic imaging is a promising tool for several biomedical applications. For this method to be quantitative, it relies on an accurate method of the light fluence in the tissue. This is difficult deep in heterogeneous tissue, where different wavelengths of light can experience significantly different attenuation. In this work, we developed a new deep neural network to simultaneously estimate the oxygen saturation in blood vessels and segment the vessels from the surrounding background tissue. The network was trained on estimated initial pressure distributions from three-dimensional Monte Carlo simulations of light transport in breast tissue. The network estimated vascular $SO_2$ in less than 50 ms with as little as 5.1% median error and better than 95% segmentation accuracy. Overall, these results show that the blood oxygenation can be quantitatively mapped in real-time with high accuracy.**

**Introduction.** Photoacoustic (PA) imaging (also known as optoacoustic imaging) relies on the absorption of pulsed laser light in tissue to generate broadband acoustic waves centered at the absorbers.[1-3] These acoustic waves are detected at the tissue surface with an ultrasound transducer. Images are reconstructed by using the time-of-flight of the acoustic waves to pinpoint their source. The end result is a technique which provides contrast based on the tissue optical properties, but has resolution determined by the acoustic receiver. This makes it particularly well suited to visualize vasculature centimeters deep in tissue.

Spectroscopic PA (sPA) imaging uses a tunable laser to acquire PA images with different wavelengths of light. The optical absorption of most endogenous chromophores and exogenous contrast agents vary significantly across the visible and near-infrared spectrum. Thus, the concentrations of the absorbers can be estimated from a multi-wavelength dataset by solving a system of equations to determine the contribution of each absorber to the PA signal at each wavelength. This approach has been applied to image oxyhemoglobin ($Hb_{O2}$) and deoxyhemoglobin (Hb). Then the relative concentrations of Hb and $Hb_{O2}$ can be used to estimate the blood oxygen saturation ($SO_2$) throughout the tissue.[4-7]

Although the prospect to noninvasively image tissue function is appealing, the accuracy of the technique is hampered by unknown tissue properties (i.e., optical scattering, absorption, and scattering anisotropy). The amount of light reaching deeper regions is dependent on the heterogeneous tissue composition in shallower regions. Furthermore, this attenuation varies as a function of optical wavelength, a phenomenon known as "spectral coloring".[8] Thus, the blood oxygenation measurements are skewed as a function of tissue composition and imaging depth, making quantitation difficult.

Recent efforts in quantifying blood saturation with sPA imaging have focused on principle component analysis or machine learning to counteract the effects of spectral coloring.[9-13] While the principle component analysis approach shows excellent quantitative performance, it is computationally intensive and requires precise knowledge of boundary conditions, making it difficult to generalize.[9, 10] The previous machine learning approaches, however, did not use Monte Carlo simulations of light transport[11, 12] or only used 2-D Monte Carlo simulates to generate the training data.[13] This means that although the datasets can be more rapidly synthesized, their accuracy is likely limited in deeper regions. Furthermore, previous deep learning approaches relied on sPA images spanning several optical wavelengths and/or low levels of noise, limiting their applicability for real-time imaging deep in tissue.[12, 13]

Here we present a new machine learning approach to obtain quantitative $SO_2$ from sPA data based on a deep convolutional neural network. Specifically, we have developed a new deep convolutional neural network – O-Net – that simultaneously estimates vascular $SO_2$ and segments the blood vessels from two-wavelength sPA data generated from three-dimensional Monte Carlo simulations of light transport in tissue. The segmentation improves the $SO_2$ prediction accuracy in blood vessels while enabling clear visualization of the vasculature. The result is a method to accurately measure the $SO_2$ in deep vessels in real-time.

**Data Generation.** The simulated sPA data were generated by MCXYZ program.[14] A 3.8 x 3.8 x 3.8 cm volume was constructed with background tissue properties selected from representative epidermis, dermis, and breast tissue measurements.[15] Between one and three cylinders with diameters ranging from 0.5 and 4 mm were inserted into the volume with a random orientation. The cylinders were filled with blood with a randomly selected $SO_2$. A laser beam with a 36 mm x 1.5 mm rectangular aperture was applied normal to the skin surface. The tissue parameters were simulated using two different optical wavelengths: 700 nm and 900 nm. A total of 4000 simulations (2000 for each wavelength) were run on a cluster where each node contained two 10-core Xeon E5-2640v4 2.40GHz processors with 256 GB of memory. Each simulation was run for 30 minutes which resulted in approximately $10^6$ photon packets per simulated volume. The resulting 128x128x128-voxel absorbed energy maps were filtered by a 3x3x3-voxel median filter to reduce the effects of a discrete number of photon packets. Finally, a single two-dimensional cross section from the center of the absorbed energy volume was taken to be the reconstructed PA image. White Gaussian noise was added to the final absorbed energy mask and the signal to noise ratio (SNR) was calculated as the average PA signal inside the blood vessels and the standard deviation of the noise.

**Model construction and training.** The deep neural network was based on the widely used U-Net architecture,[16] which has been shown to be effective at biomedical image segmentation or other tasks where the output resembles the input.[17] The architecture of the O-Net used in this study (**Fig. 1.**) consists of two U-Nets arranged in parallel. The top half of the neural network is focused on the segmentation of the blood vessels while the bottom half is used to estimate the SO2 in the vessels. In general, ELU activation functions were used except for in the last stage, where a sigmoid function was applied for the $SO_2$ image. Although sigmoid activations are typically used for classification problems, it is appropriate in this case because the output $SO_2$ is bound between 0 and 1 this activation demonstrated better accuracy than linear, ELU, or RELU activations. Dropout layers were added after the first convolution layer in each level to avoid overfitting. The input consisted of the 128x128-pixel, 700- and 900-nm PA image pair (**Fig. 2**). It was assumed that the initial pressure distribution was perfectly reconstructed to focus on the optical problem. Other computational approaches have demonstrated good accuracy in reconstruction of the initial pressure distribution and could be readily combined with the O-Net.[18, 19] The top ten rows of the PA images were zeroed out to remove the dominating signal coming from the epidermis and to allow the network to focus on the blood vessels. The output was a pair of 128x128 images. The first image (corresponding to the top branch of the O-Net) was a segmentation image of the blood vessels. The second image was the $SO_2$ map throughout the tissue.

A custom loss metric was developed for this application. The mean-squared error of the $SO_2$ was minimized only in the blood vessels (as segmented by the ground-truth oxygen maps). When a standard loss function is used the $SO_2$ was optimized predominantly for the background breast tissue, rather than the blood vessels. While this approach necessarily leads to large errors in the background tissue oxygenation estimates, these are neglected by the accompanying vessel segmentation map. A simple mean-squared error metric was used for the segmentation branch. The two losses ($SO_2$ and segmentation) were weighted equally. 80%, 10%, and 10% of the data was randomly sorted into training, validation, and testing datasets, respectively. Each image pair was normalized by the maximum pixel value to result in a range of 0 to 1. An Adam optimizer was used with a learning rate of 0.001. The networks were trained with a batch size of 32 for a total of 60 epochs. All training was performed on an Intel i5-8350U CPU with 16 GB of RAM. Training time was approximately 4.3 hours.

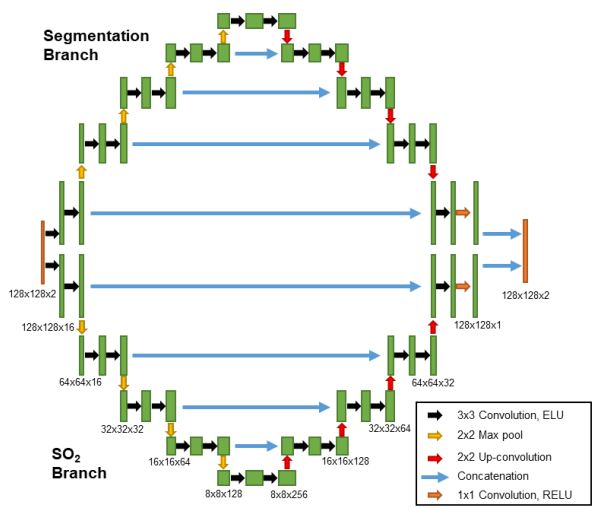

Fig. 1. O-Net network architecture used for blood vessel segmentation and oxygen estimation.

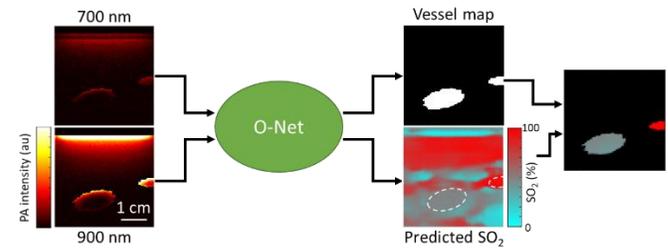

Fig. 2. Representative inputs and outputs for the O-Net. The vessel segmentation map is applied as a threshold for $SO_2$ display.

**Linear spectral unmixing.** The tissue was assumed to contain only two optical absorbers – $Hb_{O2}$ and Hb – whose absorption spectrum is known.[15] Then, the PA signal in each pixel was assumed to be proportional to the weighted sum of the two components:

$$PA(\lambda) = \mu_{a_{Hb_{O2}}}(\lambda)\, C[Hb_{O2}] + \mu_{a_{Hb}}(\lambda)\, C[Hb], \quad (1)$$

where $\lambda$ is the optical wavelength, $\mu_a$ is the molar optical absorption coefficient, and $C$ denotes the concentration. After two PA images are acquired using independent wavelengths, **Eq. 1** becomes a system of two equations and two unknowns ($C[Hb]$ and $C[Hb_{O2}]$).

We solved these equations by inverting a matrix containing the $\mu_a$ values for each wavelength. Then, the $SO_2$ in each pixel was calculated as:

$$SO_2 = \frac{C[Hb_{O2}]}{C[Hb_{O2}] + C[Hb]}. \quad (2)$$

**$SO_2$ prediction results.** The O-Net was first trained on simulated images with 25-dB average SNR. Representative images are shown in **Fig. 3**. The results from linear spectral unmixing are included as a reference. In general, the proposed O-Net yield robust results in terms of vessel size, location, and orientation. The segmentation branch was able to accurately delineate blood vessels at this level of noise. The minor errors in segmentation (e.g., **Fig. 3 b ii**) occurred in large, deep vessels where the PA signal did not persist throughout the entirety of the vessel.

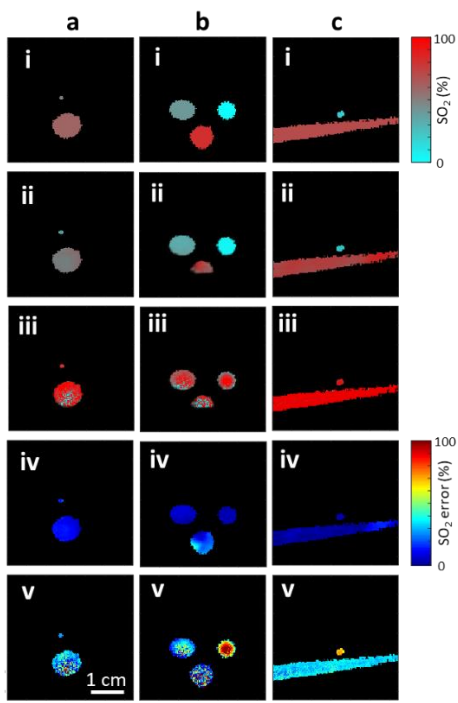

Fig. 3. (a-c) Representative results from simulations with 25 dB SNR. The panels show (i) the segmented ground truth, (ii) O-NET output, (iii) linear spectral unmixing segmented with the O-NET output, (iv) Error in $SO_2$ in the O-NET images, and (v) Error in $SO_2$ in the linear spectral unmixing images.

In general, the linear spectral unmixing method tended to over-predict the blood $SO_2$. This is because the 900-nm light is able to more readily penetrate the superficial tissue layers, which have lower absorption and scattering at longer wavelengths. Thus, the PA signal at 900 nm is greater, resulting in an over-estimation of $C[Hb_{O2}]$. In addition, the added noise (which has a greater effect in the middle and bottom of blood vessels where the PA signal is lowest) results in unpredictable estimates. Therefore, even if the general trends of spectral coloring could be corrected for (e.g., by estimating the local fluence[20]), then the noisy PA data would still lead to errors. Given that PA imaging typically suffers from low SNR in deep tissue, this ultimately limits the applicability of the linear spectral unmixing approach.

**Robustness to noise.** In order to test the O-NET's performance in the presence of larger levels of noise, the added Gaussian noise was tuned to result in an average image SNR of 5, 10, 15, 20, or 25 dB. The O-NET was then retrained for each noise level. The performance of the neural network on $SO_2$ estimation within the blood vessels of the testing dataset for each noise level is shown in **Fig. 4a**. There was little change in the performance of the linear spectral unmixing case across all noise levels. This is likely due to the fact that there was a high baseline error arising from the spectral coloring phenomenon. Thus, the addition of noise resulted in more variability of the $SO_2$ estimates within each vessel while the overall error remained consistent. The O-NET yielded much more accurate results across noise levels. As expected, the performance did degrade as a function of increasing noise level (from a median pixel-wise absolute error of 5.1% for 25-dB SNR to 13.0% for 5-dB SNR).

The O-NET segmentation performance was also inversely related to the average image SNR (**Fig. 4b**). Because the majority of the pixels correspond to the background, the O-NET tended to generate a greater number of false-negative pixels than false-positive pixels. As the noise level increased, the deeper blood vessels were less distinguishable from the background, resulting in the rise in the false negative rate (from 4.0% at 25-dB SNR to 19.6% at 5 dB SNR).

There are a few approaches which could be applied to reduce the noise and also improve the performance of the O-NET. First, acquiring multiple PA images at each wavelength and averaging the results would suppress uncorrelated noise and improve the $SO_2$ estimation. Second, images from additional wavelengths could be included into the network to enable more robust measurements. Third, the precise wavelengths used to acquire PA images could be optimized to promote the discrimination between Hb and $Hb_{O2}$.[21, 22]

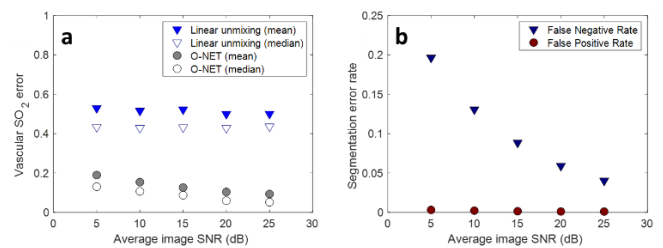

Fig. 4. (a) The mean and median absolute $SO_2$ error for all the pixels in the test dataset for linear spectral unmixing (triangles) and O-NET (circles) and (b) the segmentation false negative rate (blue triangles) and false positive rate (red circles).

**Discussion.** Overall, this proposed O-NET neural network has proven effective at estimating $SO_2$ in simulated blood vessels at greater than 3 cm deep in tissue. The largest errors in $SO_2$ manifested in large vessels in close proximity to other vessels with different $SO_2$. This phenomenon is likely to not be a major factor in clinical imaging, where PA imaging is often applied to smaller, more homogeneous vessels (e.g., for hypoxia detection in cancer applications[1, 4, 23]).

One of the biggest challenges to applying deep learning to this problem is the generation of a large dataset. A Monte Carlo simulation is the most commonly accepted method to accurately estimate light propagation in heterogeneous tissue. The approach, however is very computationally intensive, relying on stochastic modeling of millions of photon packets. Thus, building a large dataset is difficult. Training the model on a subset of our data (**Fig. 5**) indicates that additional training data will not significantly improve the performance of the neural network.

One of the key benefits of the proposed O-NET is its computational speed. In fact, the PA image pair can be segmented and the $SO_2$ can be estimated in less than 50 ms on a laptop computer. Therefore, the method could be implemented for real-time visualization of vascular oxygenation. We are currently working towards implementing it on a custom clinical PA imaging system which will be applied to imaging vascular $SO_2$ in the lymph nodes of breast cancer patients to detect metastases.[4]

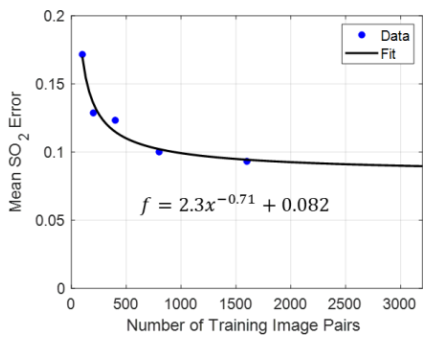

Fig. 5. Performance of the O-NET vascular $SO_2$ estimation on images with 25-dB SNR when using subsets of the training data (validation and testing data were kept constant). The power-law fit indicates that a reduction of up to 1% of $SO_2$ error is possible if the training dataset is expanded.

In order to apply the O-NET to real PA data, some additional data generation and training would be required. While Monte Carlo simulations provide a good estimate of light propagation, it is likely that a real imaging system could skew the results (for example, if the light source did not match the assumed source in the simulations). Thus, a transfer learning approach could be applied to adapt the O-NET to a real imaging system.[24, 25] In general, transfer learning requires much smaller datasets than the initial training. We expect that it would be reasonable to acquire a smaller number (on the scale of hundreds) of images of phantoms with blood-filled inclusions with well-controlled $SO_2$.

**Conclusions.** We have developed a new neural network that simultaneously estimates blood $SO_2$ and segments blood vessels from a 700-nm and 900-nm PA image pair. The network vastly outperforms linear spectral unmixing on a dataset generated from 3-D Monte Carlo simulations. Overall, this approach could be used to implement quantitative oximetry deep in tissue in real-time.

**Funding.** Neukom Institute for Computational Science, Dartmouth College.

**Disclosures**. The authors declare no conflicts of interest.